\documentclass[reprint,english,aps, prl, twocolumn, amsmath, amssymb, showpacs, showkeys, superscriptaddress]{revtex4}
\usepackage[T1]{fontenc}
\usepackage{pdfpages}
\usepackage{subfigure}
\usepackage[latin1]{inputenc}
\usepackage{graphicx}
\usepackage{epsfig}
\usepackage{prettyref}
\usepackage{babel}
\usepackage{caption}
\makeatother

\begin{document}

\title{Observation of quantum Hall effect in a microstrained Bi$_2$Se$_3$ single crystal}

\author{Devendra Kumar}
\email{deveniit@gmail.com}
\affiliation{UGC-DAE Consortium for Scientific Research, University Campus, Khandwa Road, Indore-452001, India}

\author{Archana Lakhani}
\email{archnalakhani@gmail.com}
\affiliation{UGC-DAE Consortium for Scientific Research, University Campus, Khandwa Road, Indore-452001, India}

\begin{abstract}

We report the observation of quantum Hall effect (QHE) in a Bi$_2$Se$_3$ single crystal having  carrier concentration ($n$) $\sim1.13\times10^{19}$cm$^{-3}$, three dimensional Fermi surface and bulk transport characteristics. The plateaus in Hall resistivity coincide with minima of Shubnikov de Haas oscillations in resistivity.
Our results demonstrate that the presence of perfect two dimensional transport is not an essential condition for QHE in Bi$_2$Se$_3$.
The results of high resolution x-ray diffraction (HRXRD), energy-dispersive x-ray spectroscopy (EDX), and residual resistivity measurements show the presence of enhanced crystalline defects and microstrain. We propose that the formation of localized state at the edge of each Landau level due to resonance between the bulk and defect band of Bi$_2$Se$_3$ causes the quantum Hall effect.
\end{abstract}


\maketitle
\section{Introduction}
The topological insulators have drawn a lot of attention because of interesting physical properties and possible applications ranging from areas such as in spintronics, thermoelectrics, optoelectronics, and quantum computing. These materials are conducting at surface and insulating at bulk. The conducting surface states have two dimensional massless, spin-polarized Dirac electrons with suppressed backscattering.  The special behavior of surface states in topological insulator arises from the topology of electronic band structure and time reversal symmetry~\cite{Hasan}. The chalcogenide Bi$_2$Se$_3$  is a narrow gap semiconductor having topological band gap of  0.3~eV~\cite{Xia}.  The surface states of Bi$_2$Se$_3$ have been investigated by angle resolved photoemission spectroscopy (ARPES)~\cite{Xia}, scanning tunneling microscopy~\cite{Cheng}, and   Shubnikov de Haas (SdH) oscillations~\cite{Analytis1}. Normally, the Bi$_2$Se$_3$ crystals are n-type because of electron doping due to defects and dopants. The as grown crystals of Bi$_2$Se$_3$  are selenium deficient due to volatile nature of selenium, and it has been suggested that the selenium vacancies are responsible for the electron doping of these systems~\cite{Analytis1,Franz,Butch,Analytis}.  The Hall resistivity exhibits quantum oscillations for $n<\sim5\times10^{18}$cm$^{-3}$~\cite{Hor,Eto}, while for $n\sim5\times10^{19}$cm$^{-3}$ a bulk quantum Hall effect (QHE) with two  dimensional (2D)-like transport behaviour is reported~\cite{Cao1,Cao}.
The conditions of crystal growth and the resultant crystal defects play a significant role in determining the physical properties of Bi$_2$Se$_3$. The theoretical calculations of Liu et al.~\cite{Liu1} and Young et al.~\cite{Young1} suggest that the macrostrain can transform the Bi$_2$Se$_3$ system from topological to trivial insulator.  Recently, in thin films as well as in single crystal of Bi$_2$Se$_3$, it has been observed that even microstrain can cause a topological phase transition~\cite{Liu,Kumar}. In the light of above results, the study of crystal defects and their correlation with physical properties of Bi$_2$Se$_3$ is important. 
The crystal defects along with the carrier concentration can tune the electronic properties of Bi$_2$Se$_3$ from two to three dimensional and from Dirac to Schr\"odinger like systems presenting a rich avenue for possible applications in novel electronic devices.   

In this paper, we present the observation of bulk quantum Hall effect in a Bi$_2$Se$_3$ crystal and correlate its existence with crystalline imperfection and carrier concentration. Our analysis reveals that QHE can appear in a bulk Bi$_2$Se$_3$ crystal having three dimensional (3D) transport characteristic.  

\section{Experimental}
\begin{figure*}[]
\begin{centering}
\includegraphics[width=1.0\textwidth,height=8.8cm]{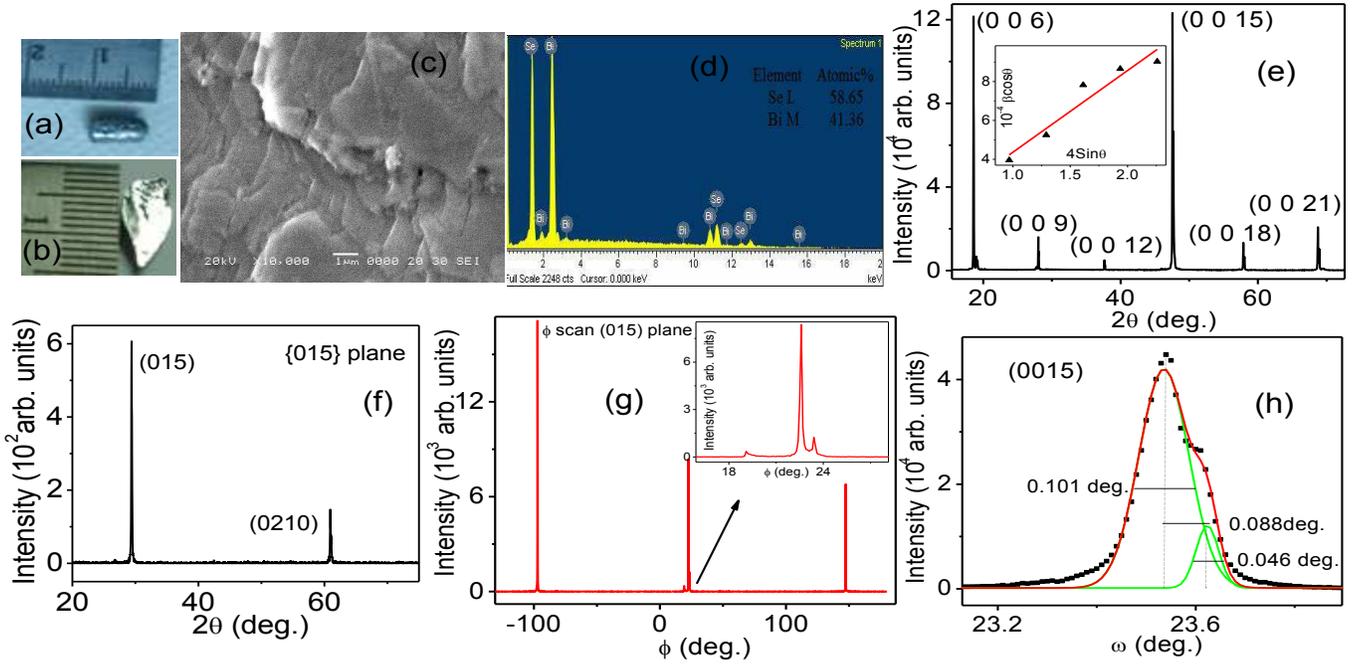}
\par\end{centering}
\caption{(a) Optical image of as grown and (b) cleaved Bi$_2$Se$_3$ crystal. (c) SEM micrograph, (d) EDX spectrum and (e) $\theta-2\theta$ x-ray diffraction pattern of the cleaved surface. The inset shows the Willamson Hall plot of Bragg peaks. The $\theta-2\theta$ x-ray data and Willamson Hall plot has been reproduced from Ref.~\cite{Kumar}. (f) $\theta-2\theta$ x-ray diffraction pattern of \{015\} plane.  (g) The azimuthal $\phi$ scan of the (015) plane. The inset shows the expanded view of peak around 22$^{\circ}$ . (h) Rocking curve for (0015) diffraction plane. The rocking curve can be deconvoluted into two curves with peaks around  23.54$^{\circ}$ and 23.62$^{\circ}$.}
\label{fig:char}
\end{figure*}

The single crystal of Bi$_2$Se$_3$ has been grown by Bridgeman method. The high purity elements of bismuth and selenium are thoroughly mixed, melted, and heated above 800$^{\circ}$C for few hours with intermediate shakes to improve the homogeneity of molten state. The crystal growth is done by slowly moving the melt through a controlled temperature gradient. The as grown crystal of Bi$_2$Se$_3$ (Fig.\ref{fig:char}~(a)) is cleaved which yields a smooth and shiny surface (Fig.\ref{fig:char}~(b)). The cleaved crystal is characterized through HRXRD, SEM and EDX measurements. For resistivity and Hall  measurements, indium contacts are made on a freshly cleaved surface and the measurements are carried on a 9T AC Transport PPMS (Quantum Design) system. The resistivity measurements are performed by conventional four probe method whereas five probe method is used for Hall measurements in order to omit the contribution of longitudinal resistivity.

\section{Results and Discussion}
Figure~\ref{fig:char}~(c) displays the SEM micrograph of the cleaved surface.  The SEM micrograph does not show any signature of crystallites at the cleaved surface indicating the single crystal nature of the sample. The EDX spectrum recorded at this surface is shown in Fig.~\ref{fig:char}~(d). The spectrum confirms the presence of only Bi and Se elements with atomic ratio of Bi:Se$\sim$2:2.8, indicating the presence of selenium vacancies.  Figure~\ref{fig:char}~(e) shows the $\theta$-2$\theta$ x-ray diffraction of the cleaved surface of our crystal. The x-ray data exhibits peaks only corresponding to \{003\} planes suggesting that the crystal is highly $<$003$>$ oriented, and the cleaved surface is perpendicular to $<$003$>$ (C$_3$) axis of Bi$_2$Se$_3$. Figure~\ref{fig:char}~(f) exhibits $\theta$-2$\theta$ x-ray diffraction data for \{015\} planes. The data have peaks only corresponding to \{015\} planes indicating the single crystalline nature of our sample.

Figure~\ref{fig:char}~(g) displays the azimuthal scan of (015) diffraction plane.  The scan shows three peaks corresponding to three fold symmetry of (015) plane demonstrating the in plane orientation.  A close observation of the peak around 22$^{\circ}$ shows the presence of two tiny peaks in the vicinity of main peak. This indicates the presence of small disorder in the in-plane orientation of some regions of sample. These crystalline defects along with selenium vacancies can create the local strain field in the crystal. The strength of microstrain ($\epsilon$) in a crystalline material can be investigated through Willamson-Hall relation $\beta$Cos$\theta$ = $K\lambda/D$ + (4$\epsilon$Sin$\theta$), where $\theta$ is the angle of Bragg peak, $\beta$ is the full width at half maximum (FWHM) of Bragg peak, $K$ is the Scherrer constant, $\lambda$ is the wavelength of x-ray, and $D$ is the crystallite size~\cite{Emil}. The Willamson Hall analysis of the Bragg peaks of \{003\} planes gives $\epsilon\sim$~4$\times10^{-4}$~\cite{Kumar}. See inset of Fig.~\ref{fig:char}~(a).  Strain $\epsilon\sim$~10$^{-6}$ is observed in nearly perfect single crystals of silicon~\cite{Kushwaha}. The large value of $\epsilon$ in our crystal is a clear indication of enhanced local strain field. The positive sign of $\epsilon$ suggests that the major contribution to microstrain is not from selenium vacancies, as the vacancies cause local tensile strain which  yields a negative $\epsilon$~\cite{Kushwaha}. The positive $\epsilon$ signals the presence of dominant local compressive strain field possibly arising from the lattice disorder other than Se vacancies.

Figure~\ref{fig:char}~(h) shows the rocking curve for (0015) plane of our crystal. The rocking curve measurement is an important tool to investigate the crystalline perfection of the crystal. The rocking curve peak has intensities over a broad angular region (~0.5 degree) and contains a shoulder on the higher $\omega$ side. The deconvolution of rocking curve gives two peaks at a separation of 0.088 degree. A second peak in the rocking curve suggests the presence of low angle grain boundary  due to presence of slight misorientation in crystal domains~\cite{Kushwaha}. The FWHM of the main peak of rocking curve is $\sim$0.101 degree. The observed FWHM of our crystal is three times larger in comparison to a nearly perfect strain free silicon crystal (~0.03 deg. measured on the same machine).  The relatively large FWHM of our rocking curve suggests the presence of enhanced crystalline disorder such as dislocation, mosaic spread, and misorientation.

\begin{figure}[]
\begin{centering}
\includegraphics[width=1.0\columnwidth]{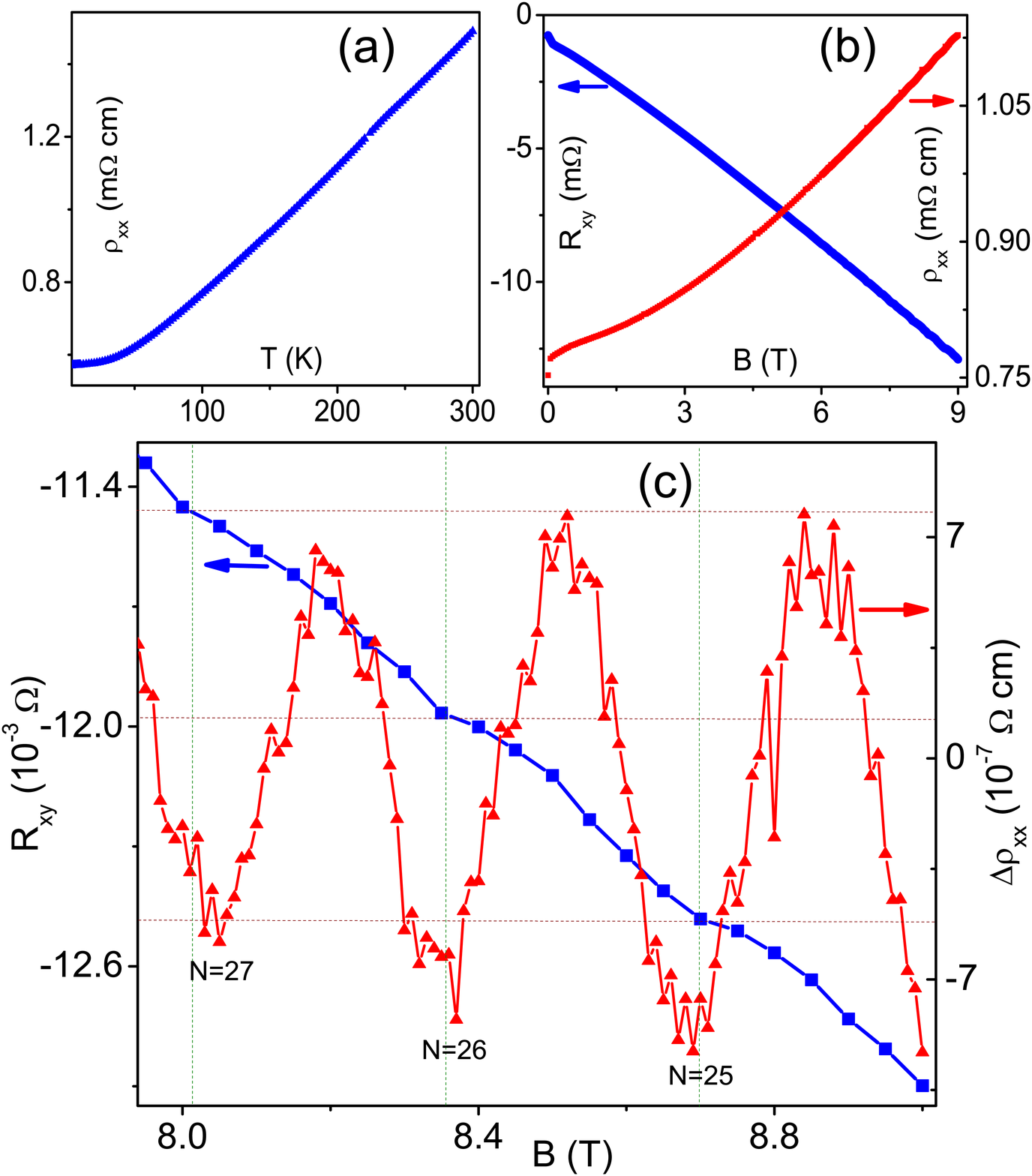}
\par\end{centering}
\caption{(a) Temperature dependence of longitudinal resistivity ($\rho_{xx}$).  (b) Hall resistance (R$_{xy}$) and longitudinal resistivity ($\rho_{xx}$) as a function of magnetic field ($B||C_3$ axis) at 1.8~K. The Hall and resistivity data has been reproduced from Ref.~\cite{Kumar}.(c) Expanded view of the Hall resistance along with oscillatory component of magnetoresistance ($\Delta\rho_{xx}$) above 8~T at 1.8~K.}
\label{fig:Hall}
\end{figure}

Figure~\ref{fig:Hall}~(a) shows the temperature dependence of resistivity of our Bi$_2$Se$_3$ crystal.  The low temperature residual resistivity $\rho_0$ of our sample is  $\sim$0.574m$\Omega$ cm. Figure~\ref{fig:Hall}~(b) exhibits the results of Hall resistivity and magnetoresistance at 1.8~K. The Hall resistivity gives carrier concentration $n\sim$1.13$\times$10$^{19}$cm$^{-3}$~\cite{Kumar}.  A comparison of residual resistivity of our crystal with the samples of similar carrier concentration ($n\sim10^{19}$cm$^{-3}$) from literature (0.125m$\Omega$~cm~\cite{Butch}, 0.26m$\Omega$~cm~\cite{Analytis}, 0.330m$\Omega$~cm~\cite{Cao1}) shows that our crystal has a larger residual resistivity. The low temperature residual resistivity represents the contribution of defect/impurity scattering, and therefore, the larger residual resistivity of our sample is an outcome of higher defect density of our crystal. The magnetoresistance data at 1.8~K exhibits SdH oscillations above 6~T and the analysis of these oscillations give Dingle temperature (T$_D$) $\sim$ 40~K~\cite{Kumar}. The Dingle temperature accounts for the electron scattering and the large value of T$_D$  in our sample further corroborates the presence of enhanced crystalline defects~\cite{Shoenberg}.

Figure~\ref{fig:Hall}~(c) displays the Hall resistance along with the SdH oscillations in magnetoresistance above 8~T at 1.8~K. The Hall resistance above 8~T exhibits plateaus coinciding with the minima of resistivity oscillations. The presence of plateaus in the Hall resistance at the minima positions of SdH oscillations is a signature of Quantum Hall effect. The Quantum Hall effect in our system is from the bulk sample. Similar observation of quantum Hall effect in the bulk sample of Bi$_2$Se$_3$ has been reported by Cao et al.~\cite{Cao1,Cao}. Their results show that the bulk quantum Hall resistance ($R_{xy}$) in Bi$_2$Se$_3$ takes value of $R_{xy}\sim\frac{1}{NZ}\frac{h}{e^2}$, where $N$ is the Landau level index and $Z$ is the thickness of the sample in the units of quintuple layer ($\sim$~1nm). These results were interpreted as the contribution of many parallel 2D conduction channels, where each conduction channel is a quintuple layer of Bi$_2$Se$_3$. The phenomenon of quantum Hall effect is not limited to perfectly two dimensional systems and is reported to occur in a number of bulk systems with high anisotropy and layered structures, 
such as GaAs/AlGaAs multi-quantum wells~\cite{Baldwin}, graphite~\cite{Kempa,Bernevig}, Bechgaard salts~\cite{Hannahs,Balicas}, Mo$_4$O$_{11}$~\cite{Hill}, Bi$_{2-x}$Sn$_x$Te$_3$~\cite{Inouea}, and Sb$_{2-x}$Sn$_x$Te$_3$~\cite{Sasaki}. The bulk quantum Hall effect has been associated with the formation of charge or spin density waves~\cite{Hannahs,Balicas,Hill}, weak inter-layer coupling~\cite{Baldwin,Kempa,Bernevig}, or the charge transfer from lower valence or impurity band to bulk Landau levels~\cite{Inouea,Sasaki}.  The expected quantized Hall resistance from the contribution of parallel quintuple layers for our sample (sample thickness $\sim$0.3~mm, $\sim3\times$10$^5$ quintuple layers) is $\sim$ ~3.19~m$\Omega$ for N=27, 3.31~m$\Omega$ for  N=26, 3.44~m$\Omega$ for  N=25. We get the plateaus in Hall resistance at 11.4508~m$\Omega$ (for N=27), 11.9664~m$\Omega$ (for N=26) and 12.4817~m$\Omega$ (for N=25).  The observed Hall resistance is of the order of expected values from parallel quintuple layers. In quantum Hall effect, the ratios of Hall resistance should match with the inverse of corresponding ratios of Landau Levels ($N$). These ratios match well for our sample  e.g. $R_{xy}(N=25)/R_{xy}(N=27)$ $\sim$ 1.09 matches with $27/25(1.08)$ supporting the quantized nature of Hall resistivity.

The previous report of quantum Hall effect in Bi$_2$Se$_3$ (by Cao et al.~\cite{Cao1}) used a highly doped sample $n\sim4.7\times10^{19}$~cm$^{-3}$, and suggested that at $n\geq3\times10^{19}$~cm$^{-3}$, the transport behavior of Bi$_2$Se$_3$ changes from bulk to  stacked 2D like conduction channels. The appearance of 2D like conduction channels at high $n$ needs to be substantiated by other experimental or theoretical reports.   For our sample, the analysis of Landau level fan diagram and the angle dependence of SdH oscillations show the presence of Dirac electrons with anisotropic 3D Fermi surface and bulk transport characteristic~\cite{Kumar}. The observation of QHE in the bulk transport regime of Bi$_2$Se$_3$ demonstrates that a transition from 3D to perfect 2D transport is not an essential requirement for QHE in Bi$_2$Se$_3$. 
Moreover, a comparison of our results with that of Cao et al.~\cite{Cao1,Cao} suggest that if there is an $n$ dependent crossover from 3D to 2D transport, it should occur between $n\sim$1.13$\times$10$^{19}$ to $\sim4.7\times10^{19}$~cm$^{-3}$.

While the exact physical reason for existence of quantum Hall effect in 3D topological insulator is still an open question, Cao et al.~\cite{Cao1} suggested that possibly the enhanced warping and nesting of bulk conduction band at high $n$ drives the formation of charge density waves which are important for observation of bulk quantum Hall effect~\cite{Hill}. The hexagonal wrapping of Fermi surface of surface states at high carrier concentration has been observed in ARPES studies~\cite{Kuroda} but no such deformation of bulk states has been reported till now. Moreover, the temperature dependence of resistivity (see Fig.~\ref{fig:Hall}~(a)) does not show any anomaly signalling the absence of charge density wave formation in the temperature range of measurements. The bulk quantum Hall effect can also occur in narrow band semiconductors having disorder broadened defects/impurity band near the Fermi energy and 3D Landau levels in bulk band. At high fields, the resonance between the impurity band and the bulk band forms the localized state at the edge of each Landau level and these localized states cause the quantum Hall effect in these systems~\cite{Inouea,Sasaki,Mani}. Such 3D quantum Hall effect has been observed in Bi$_{2-x}$Sn$_x$Te$_3$~\cite{Inouea}, Sb$_{2-x}$Sn$_x$Te$_3$~\cite{Sasaki}, Hg$_{1-x}$Cd$_x$Te~\cite{Mani}, and InSb~\cite{Mani}.
Similar quantum Hall effect has been also observed in Bi$_2$Te$_3$ where it is attributed to the charge transfer from the Landau levels of lower valance band to upper valance band~\cite{Inouea}.

Our sample has carrier concentration $n\sim1.13\times$10$^{19}$~cm$^{-3}$, smaller than Cao et al. ($n\sim4.7\times10^{19}$~cm$^{-3}$)~\cite{Cao1,Cao}, but exhibits quantum Hall effect at low temperatures. The observation of quantum Hall effect in samples having $n\ge10^{19}$~cm$^{-3}$ but its absence in samples with $n\leq5\times10^{18}$cm$^{-3}$~\cite{Hor,Eto} suggest that if QHE is only governed by carrier concentration ($n$), the critical concentration for QHE lies between $n\sim5\times10^{18}$ to $1\times10^{19}$cm$^{-3}$.
The other important characteristic of Bi$_2$Se$_3$ samples exhibiting QHE is their relatively large Dingle temperature (T$_D$); T$_D$ for our sample is $\sim$ 40~K~\cite{Kumar}, for Cao et al. T$_D$ is $\sim$ 25~K~\cite{Cao1,Cao}, while for samples without QHE T$_D$ is $\le$10~K~\cite{Hor,Eto}. The large Dingle temperature signals the presence of higher crystalline defects, which is indeed the case for our sample as revealed by HRXRD and high residual resistivity. Recent hard x-ray photoemission experiments show the presence of vacancies/defects states close to the Fermi energy of Bi$_2$Se$_3$~\cite{Biswas}. The resonance between the disorder broadened vacancies/defects band and the 3D Landau levels could cause the quantum Hall effect in our system.  The appearance of quantum Hall effect in highly doped and disordered samples (present study, Ref.~\cite{Cao1}\&\cite{Cao}) and its absence in less doped and clean samples (Ref.~\cite{Hor}\&\cite{Eto}) supports this conjuncture.
Further systematic investigations on samples with different disorder by varying $n$ either by substitution or by creating vacancies would help in determining whether the charge carrier density $n$ or the crystalline defects/disorder dictates the appearance of QHE and the  physical mechanism responsible for it.

\section{Conclusion}
In conclusion, we present the observation of quantum Hall effect (QHE) in a Bi$_2$Se$_3$ crystal, and discusses its possible origin. Our results show that QHE in our system is not from 2D surface states but from the bulk of Bi$_2$Se$_3$ having 3D Fermi surface. These results are important, as they reveal that the perfect 2D transport is not an essential criterion for QHE in Bi$_2$Se$_3$. The bulk QHE in Bi$_2$Se$_3$ is ascribed to formation of localized state at the edge of Landau levels due to presence of disorder broadened defect band near Fermi energy. 
A comparison of our results with literature suggest that a transition from QHE to non-QHE behavior is expected between n$\sim$1$\times$10$^{19}$ cm$^{-3}$ to 5$\times$10$^{18}$ cm$^{-3}$. The correlation of QHE with the carrier concentration and microstrain indicate that Bi$_2$Se$_3$ can be tuned in its various transport regimes e.g  QHE to non-QHE, 2D to 3D and Dirac to non-Dirac conduction on variation of carrier concentration and strain/micro-strain.



\section{Acknowledgements}
We thank V. R. Reddy for HRXRD measurements and D. M. Phase for SEM and EDX measurements.  V. Ganesan and A. K. Sinha are acknowledged for support and encouragement.

\end{document}